# Highly Sensitive On-Chip Magnetometer with Saturable Absorbers in Two-Color Microcavities

O. Gazzano[1,*,†] and C. Becher[1]

[1] Fachrichtung 7.2 (Experimentalphysik), Universität des Saarlandes, Campus E2.6, 66123 Saarbrücken, Germany
*Corresponding author: ogazzano@umd.edu
†Present address: Joint Quantum Institute, National Institute of Standards and Technology, & University of Maryland, Gaithersburg, MD, USA.

Interacting resonators can lead to strong non-linearities but the details can be complicated to predict. In this work, we study the non-linearities introduced by two nested microcavities that interact with nitrogen vacancy centers in a diamond waveguide. Each cavity has differently designed resonance; one in the green and one in the infrared. The magnetic-field dependence of the nitrogen vacancy center absorption rates on the green and the recently observed infrared transitions allows us to propose a scalable on-chip magnetometer that combines high magnetic-field sensitivity and micrometer spatial resolution. By investigating the system behaviors over several intrinsic and extrinsic parameters, we explain the main non-linearities induced by the NV centers and enhanced by the cavities. We finally show that the cavities can improve the magnetic-field sensitivity by up to two orders of magnitudes.

## 1. INTRODUCTION

Nitrogen vacancy (NV) centers in diamond can measure weak magnetic fields with nanoscale resolution, and thus have aroused broad interests for magnetic sensing applications [1–3]. These applications cover many diverse fields such as imaging in neuroscience, biology or microfluidics [4–6], and for controlling domain walls of magnetic devices [7]. The sensitivity can be improved by a factor of about 3 by increasing the source brightness by shaping the surrounding diamond [8–11]. Cavity electrodynamic effects could lead to larger brightnesses [12] however, the required Purcell enhancement of the emission rate has only be observed at cryogenic temperatures because the zero-phonon line of a NV center is broadened with the temperature [13–15]. Furthermore, devices require a compromise between the magnetic-field sensitivity and the spatial resolution, since to increase sensitivity the size of the active region has to be enlarged to include more NV centers [3, 16].

An alternative solution is to use the infrared transition that has been recently observed in the singlet states of NV centers (Fig. 1a, [17]). Although its fluorescence emission rate is weak, the microsecond lifetime of its ground state $|6\rangle$ allows for absorption measurements [18–20]. Moreover, since the decay rates from the excited electronic spin states $|3\rangle$ and $|4\rangle$ to the singlet level $|5\rangle$ are different, a measurement of the absorption rate of infrared light can be used to perform electron spin resonance (ESR) spectroscopy and to deduce the applied magnetic field [21]. Indeed, for low magnetic fields, the microwave transition energy depends on the longitudinal compo-nent $B$ of the field through to the Zeeman effect [22]: $D \pm \gamma B/2\pi$ ($\gamma = 1.761 \times 10^{11}$ rad s$^{-1}$T$^{-1}$ is the gyromagnetic ratio of the electronic spin and $D = 2.87$ GHz). The ESR contrast – and therefore the sensitivity – can be improved by a resonant cavity that will increase the absorption optical path length [23, 24].

## 2. SCALABLE ON-CHIP MAGNETOMETER

### A. Proposed device

In this letter, we propose a scalable on-chip magnetometer comprising a doubly resonant microcavity system for the green pump light ($\lambda_{\mathrm{Gr}} = 532$nm) and for the infrared transition ($\lambda_{\mathrm{IR}} = 1042$nm). The cavities allow for a strong enhancement of the magnetic-field sensitivity without compromising spatial resolution since the cavities increase the absorption path length by a factor proportional to their finesse. The structure also allows for precise control, for both colors, of the intracavity fields and of the overall losses.

The magnetometer operates by sending green and infrared light to the sensor and by analyzing the intensity of the reflected or transmitted infrared beams while performing ESR spectra (Fig 1b). The sensor is a micrometer square photonic waveguide comprising two nested cavities along the axis of the waveguide (Fig. 1c). The nested cavities are formed from two sets of distributed Bragg reflectors (DBRs), and contain a high density of NV centers. The two pairs of DBRs are designed for central wavelengths at $\lambda_{\mathrm{Gr}}$ and $\lambda_{\mathrm{IR}}$. To minimize infrared probe-field losses, the infrared mirrors are located inside the green cavity. For the study, we consider a diamond waveguide and air/diamond

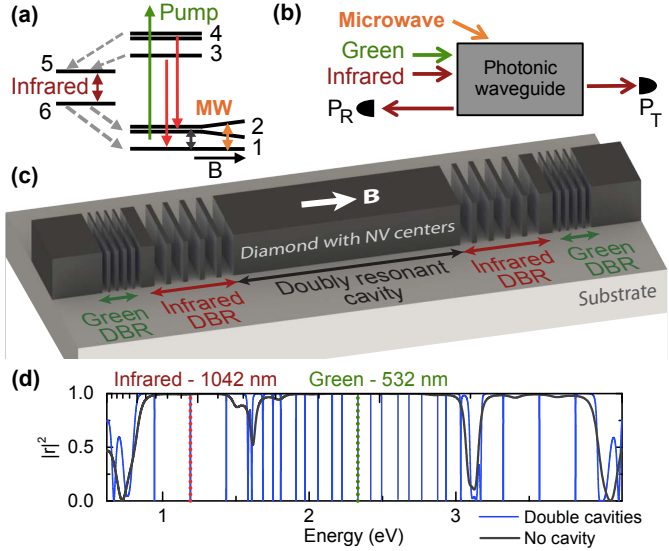

**Fig. 1.** (a) $NV^-$ center energy levels with the infrared transition. Arrows indicate the green pump (green) and the red (light red), infrared (dark red), microwave (orange) and non-radiatives (dot grey) transitions. (b) The transmitted $P_T$ or the reflected $P_R$ infrared light power can be measured. The sample is pumped with a green field. (c) The doubly resonant cavity with two pairs of 523nm and 1042nm DBRs. (d) Calculated reflectivity of two color DBRs (black) and of the two-cavity system (blue).

DBRs pairs that can be obtained by etching a diamond film [25, 26]. We use a traditional transfer matrix calculation method to determine the reflectivity of the device [27]. The full cavity features two stop-bands centered around $\lambda_{Gr}$ and $\lambda_{IR}$ and a series of Fabry Perot interferences dips (Fig. 1d). Two-dimensional finite element calculations are used to prove that both the green and the infrared fields are well confined inside the cavity (See Supplemental Material [28]).

## B. Magnetic-field sensitivity: General consideration

The magnetic-field sensitivity is proportional to the ESR linewidth, $\Gamma_{MW}$ and inversely proportional to the contrast, $C$ between the output infrared powers, $P^s$ measured with the microwave *on*- ($s = $ on) or *off*-resonance ($s = $ off) between the electronic spin states $|1\rangle$ and $|2\rangle$. The shot-noise limited sensitivity, $\delta B$ also depends on $P_{max} = \max(P^{on}, P^{off})$ and on the acquisition time $t_m$ [23, 24]:

$$\delta B = \frac{\Gamma_{MW}}{\gamma C} \sqrt{\frac{hc}{P_{max} t_m \lambda_{IR}}} \text{ and } C = \frac{|P^{on} - P^{off}|}{P_{max}} \quad (1)$$

$\delta B$ can also be written as a function of the fraction, $F_i^s$ of the infrared light that is transmitted ($i = T$) or reflected ($i = R$) by the cavity:

$$\delta B_i^{-1} \propto |F_i^{on} - F_i^{off}| / \sqrt{\max(F_i^{on}, F_i^{off})}. \quad (2)$$

In order to calculate the values of $F_i^s$, we introduce the amplitude reflectivity coefficients $\rho_i = 1 - \epsilon_i$ ($\epsilon_i \ll 1$) of the front ($i = 1$) and of the back ($i = 2$) infrared mirrors. We write $\alpha^s = \alpha_0 + \alpha_{NV}^s$, the losses in amplitude of the infrared light over the cavity length. The term $\alpha_{NV}^s \ll 1$ accounts for the absorption induced by the infrared transition of the NV centers – that depends on the NV spin state – and $\alpha_0 \ll 1$ includes all other losses (scattering and absorption by the waveguide sidewalls and by other diamond defects). We also consider forward and backward propagating waves in the waveguide. Using the phase and amplitude conservation rules at the two mirrors and during the propagation in the cavity, we find:

$$F_T^s = \frac{4\epsilon_1\epsilon_2}{(\epsilon_1 + \epsilon_2 + 2\alpha^s)^2}; \; F_R^s = \left(\frac{\epsilon_1 - (\epsilon_2 + 2\alpha^s)}{\epsilon_1 + \epsilon_2 + 2\alpha^s}\right)^2 \quad (3)$$

We first plot the shot noise limited sensitivity, $\delta B$ for the transmission case as a function of the reflectivity of the input and output mirrors (Fig. 2a). Since any asymmetry on the mirrors reduces the transmitted light intensity, transmission measurements should be performed on symmetric cavities ($\epsilon = \epsilon_1 = \epsilon_2$, red line in Fig. 2a). The value of $\epsilon$ has to be carefully chosen. When the absorption rates of the cavity medium are low compare to losses induced by the mirrors ($\alpha^s \ll \epsilon$, over-coupled cavity, upper left in Fig. 2a), $\delta B^{-1} \propto 2(\alpha_{on} - \alpha_{off})/\epsilon$ meaning that the sensitivity is linearly improved with $\epsilon$. In the opposite case, when $\alpha^s \gg \epsilon$ (under-coupled cavity, bottom right in Fig. 2a), $\delta B^{-1} \propto \epsilon/A$, where $A$ is a rational function of $\alpha_{on}$ and $\alpha_{off}$. Thus, further increases of the mirror reflectivity degrade the sensitivity.

The reflectivity configuration shows different behaviors (Fig. 2b). In the symmetric cavity case, one can show that better sensitivities can be reached only if $\alpha_{NV}^{on} \gg \alpha_{NV}^{off}$. The strongly asymmetric cavity case ($\epsilon_2 \ll \epsilon_1$) allows for better sensitivi-ties. It is optimal either when $F_R^{\text{on}}(\epsilon)$ or $F_R^{\text{off}}(\epsilon)$ are minimum (two horizontal lobes in Fig. 2b). These cases correspond to the impedance matching regime when the same amount of losses are induced by the cavity medium and by the mirrors ($\epsilon_1 = a_0 + a_{\text{NV}}^s$). Between the two lobes when $F_R^{\text{on}} = F_R^{\text{off}}$, no magnetic-field can be measured ($\delta B = +\infty$) because $C = 0$.

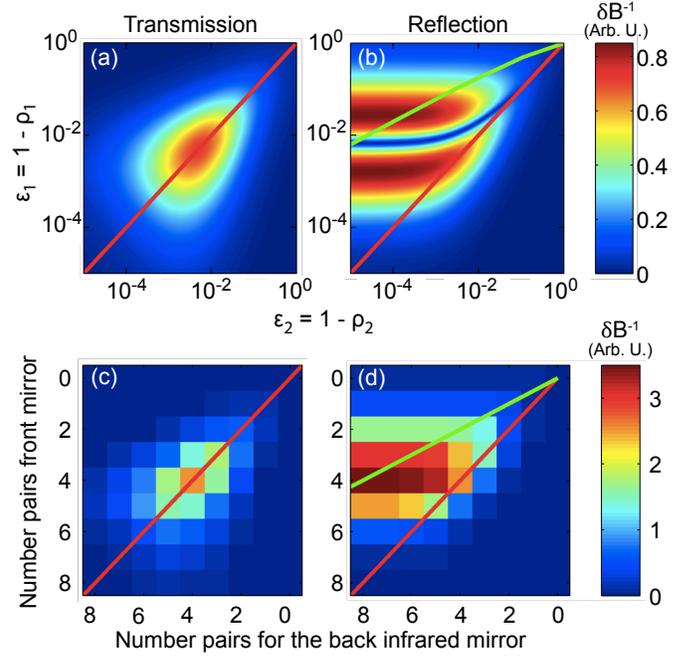

**Fig. 2.** Inverse of the magnetic-field sensitivity $\delta B^{-1}$ in transmission (a,c) and reflection (b,d) configurations: as a function of $\epsilon_1$ and $\epsilon_2$ using Eqs. 2,3 with $\alpha_0 = 10^{-3}$, $\alpha_{NV}^{on} = 10^{-2}$ and $\alpha_{NV}^{off} = 0$ (a,b) and as function of the number of pairs for the front and back infrared mirrors using the transfer matrix calculations program considering real-world parameters (c,d). The red lines correspond to a symmetric cavity and the green lines to an asymmetric cavity with twice more pairs for the back infrared mirror.

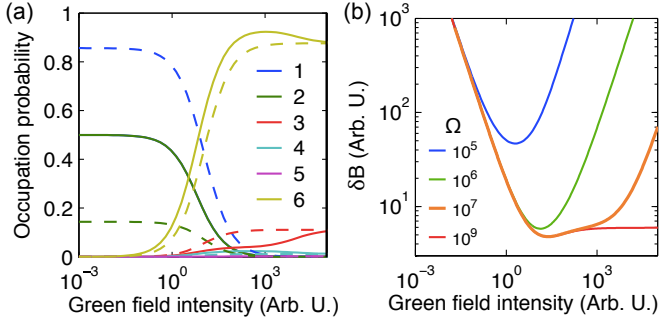

**Fig. 3.** (a) Occupation probability of the 6 energy levels as function of the green pump field intensity with the microwaves on (solid lines, $\Omega = 2\pi \times 10$ Mhz) or off (dotted lines) resonance. (b) Sensitivity to the magnetic-field for different Rabi frequencies $\Omega$ (in $2\pi$ Hz) induced by the microwave field.

We now use the rate equations model developed in Ref. [23] to calculate the occupation probabilities $N_i$ of the 6 energy levels of the NV center and to deduce the absorption rates, $a_{NV}^{\text{Gr,s}} = \sigma_{\text{Gr}}(N_1^s + N_2^s)$ and $a_{NV}^{\text{IR,s}} = \sigma_{\text{IR}}(N_6^s - N_5^s)$ of the green and infrared fields with the microwave field $on$- or $off$- resonance ($\sigma_{\text{Gr}}$ and $\sigma_{\text{IR}}$ are the absorption cross-sections [23, 29]). The model accounts for the transitions shown by arrows in Fig. 1a (See Supplemental Material [28]). As expected, the occupation probabilities of the ground levels $|1\rangle$ and $|2\rangle$ decrease when the pump field increases (Fig. 3a). In the saturation regime, the occupation probability of the infrared ground level, $|6\rangle$ is dominant due to its longer lifetime ($\simeq \mu s$). Interestingly, its occupation probability depends on the microwave field frequency – and intensity – and can therefore be used to perform ESR spectroscopies.

The relation $\delta B \propto \sqrt{P_{\max}} / \left| P^{\text{on}} - P^{\text{off}} \right|$ deduced from Eq. 1 expresses the dependence of the magnetic-field sensitivity to the infrared field power after the interaction with the NV centers. Moreover, since the microwave field allows the transitions from the electronic spin ground states $|1\rangle$ and $|2\rangle$, and the green field reinitializes the spins to $|1\rangle$, the equilibrium between these two opposite effects leads to an optimal sensitivity (minimum of the curves in Fig 3b). Stronger green and microwave fields lead to better sensitivities until the green light saturate the NV centers (red lines in Fig 3b and Fig S3a-b [28]).

## 3. EFFECT OF THE NON-LINEAR INTERACTIONS ON THE MAGNETIC-FIELD SENSITIVITY

### A. Model

The above calculations show that several parameters of the device are strongly coupled and need to be carefully chosen to optimize the sensitivity. To reveal this dependence, we use a transfer matrix calculation method to find the shot-noise limited sensitivity to the magnetic field. We account for the complete structure of the double cavity system and use real-world parameters. The medium is discretized and the absorptions rates are calculated at every computational point to account for their non-linear dependences with the local intensity of both the green and the infrared fields (See Supplemental Material [28]). We consider a $\sigma_c = (3\mu m)^2$ square cross-section diamond waveguide with $a_0^{\text{IR}} = 10 m^{-1}$. The intensity of the input infrared light is taken to equal $1 MW/m^2$ (9mW on the waveguide cross-section) and the microwave field induces a Rabi frequency of $\Omega = 2\pi \times 10 MHz$.

### B. Green cavity

This method allows us to calculate the magnetic-field sensitivity in the reflection and transmission configurations and as a function of the input green power. Since the reflection case has better sensitivity, we use it. We consider now a symmetric green cavity with $N_{\text{Gr}} \in [\![3,6]\!]$ DBR pairs, an asymmetric infrared cavity with twice more pairs on the back infrared mirror than on the front one ($N_{\text{IR}}^{\text{Front}} = 4$), a cavity length of $L_c = 120\lambda_{\text{IR}}/n_D$ ($n_D \simeq 2.4$ is the refractive index of the diamond), a density $d = 4.4 \times 10^{23} m^{-3}$ of NV centers and an electronic spin dephasing time of $T_2^* = 390ns$ according to real sample values and former calculations [21, 23].

The results, plotted in Fig. 4a., show that the dependance of the magnetic-field sensitivity on the input green power is very similar than for an isolated single NV center (orange bold curve in Fig. 3b). The sharp peaks appearing on some curves and for some pump powers correspond, once again, to equal reflectivities $F_R^{\text{on}} = F_R^{\text{off}}$ with and without the microwave field (Eq. 2). Significant improvements of the sensitivity, by more than two orders of magnitude, are found comparing with the no-cavity cases (solid and dotted black lines in Fig. 4a).

We observe in Fig. 4b, that plots the intracavity green intensity as a function of the green input power, $P_{\text{Gr}}$, that different cavity regimes can be established depending on both the green input power and on the number of green DBR pairs, $N_{Gr}$. In the considered configuration, the cavity is always over-coupled when $N_{Gr}$ is smaller than 3: the intracavity field increases with $N_{Gr} < 3$ and with the incident green power (dashed lines). However, when $N_{Gr} \geq 3$ (solid lines), the cavity regime depends on the the incident power. It remains over-coupled at high pump field – when the NV centers saturate – but is under-coupled at low pump field (strong absorption by the NV centers).

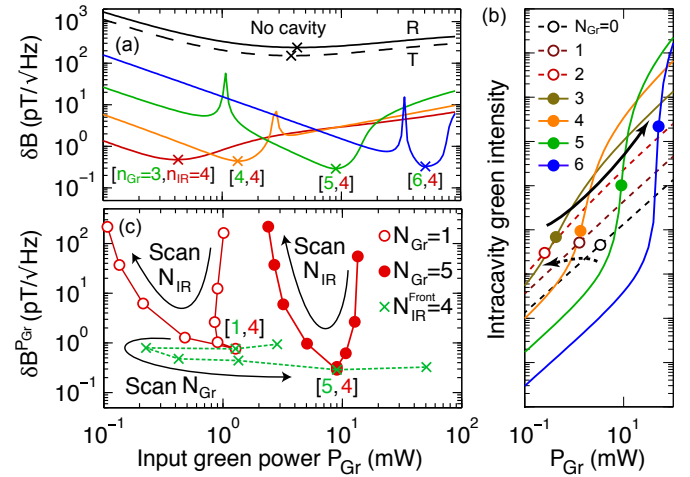

**Fig. 4.** (a) Magnetic-field sensitivity as a function of the input green power for a scan over $N_{\text{Gr}} \in [\![3,6]\!]$ in reflection (solid lines) or transmission (dashed line); numbers are $[N_{\text{Gr}}, N_{\text{IR}}^{\text{Front}}]$; the x-marks correspond to $\delta B^{P_{\text{Gr}}}$. (b) Intensity of the intracavity field extracted from the simulations runs as a function of the input power; $N_{\text{IR}}^{\text{Front}} = 4$. The marks correspond to $P_{\text{Gr}}^{\text{opti}}$. The arrows indicate the displacement of $P_{\text{Gr}}^{\text{opti}}$ when the cavity is always over-coupled (dotted lines, $N_{\text{Gr}} < 3$) or when it is under-coupled at low pump intensity (solid lines, $N_{\text{Gr}} \geq 3$). (c) Values of $\delta B^{P_{\text{Gr}}}$ for: a scan over $n_{\text{IR}}^{\text{Front}} \in [\![0,8]\!]$ with $N_{\text{Gr}} = 1$ (open circles) or $N_{\text{Gr}} = 5$ (full circles); a scan over $N_{\text{Gr}} \in [\![0,6]\!]$ with $N_{\text{IR}}^{\text{Front}} = 4$ (green x-marks).

Fig. 4b shows that the inflection points are shifted toward higher input green powers as $N_{Gr} \geq 3$ increases. The best sensitivity configuration follows the same behaviors: the optimal green input power slightly increases as $N_{Gr} \geq 3$ increases (solid marks) although it reduces when $N_{Gr} < 3$ (open marks).

These dependences of the cavity regime actually lead to an interesting effect: the slope of the sensitivity with the pump power after its change of sign (x-marks in Fig. 4a) increases with the number of green pairs $N_{Gr} \geq 3$. Indeed, the non-linearity of the green light absorption rates (it is reduced above the saturation regime of the transition) is enhanced by the cavities.

## C. Infrared cavity

In order to understand the dependance of the best magnetic-field sensitivity to the number of pairs on the front $N_{IR}^{front}$ and back $N_{IR}^{back}$ infrared DBRs, we perform simultaneous scans over the green power and over the number of green layers to find the optimum sensitivity for each pairs $[N_{IR}^{front}, N_{IR}^{back}] \in [\![0,8]\!]^2$. The graphs that we obtain (shown in Figs. 2c,d) are similar to the ones obtained with the general Eq. 2 and plotted in Figs. 2a,b. The divergence of the sensitivity that appears when $C = 0$ (Fig. 2b) is hidden by the scan over the green power because it occurs only for a precise value of $\alpha_{NV}^s$. The graphs also show that reflectivity measurements can allow for 36% better sensitivities than transmission ones.

In Fig. 4c and for the rest of the study, we consider the reflection case and asymmetric cavities with $N_{IR} = N_{IR}^{front} = N_{IR}^{back}/2$ because it allows the device to reach the maximum of sensitivity (green line in Fig. 2d). Fig. 4c shows the dependance of the optimum sensitivity $\delta B^{P_{Gr}}$ for scans over the number of pairs in the green and infrared mirrors. The sensitivity is improved by a factor of $\sim 3$ by the use of the green cavity and by a factor above 500 by the use of an infrared cavity. Fig. 4c also shows that the coupling between the two-color cavities via the NV center ensemble allows for smaller green incident powers when pairs are added on the infrared mirror. In the over-coupled regime, this behavior arises from changes in the NV center states dynamics under higher infrared field. Indeed, the infrared light shifts the optimal occupation probability of $|6\rangle$ to a lower value although the sensitivity is optimal when the state $|6\rangle$ is about half populated (See Supplemental Material [28]).

## D. Doubly resonant cavity length

We now investigate the influence of the cavity length. In Fig. 5a, we consider the configuration $[N_{Gr}, N_{IR}] = [5,4]$ and we perform green input power scans to obtain the best sensitivity $\delta B^{P_{Gr}}$ and the associated pump power $P_{Gr}^{opti}$ for cavity lengths from 10 to 400 $\lambda/n_D$ (thin line in Fig. 5a). Like the green pump dependance curves obtained above (Fig. 4a), the two local minima of the magnetic-field dependance correspond to $F_R^{on} = 0$ and $F_R^{off} = 0$ in Eq. 2 and the divergence peak to $F_R^{on} = F_R^{off}$. The curve also shows that the optimized green power increases with the cavity length (color scale of the thin line in Fig. 5a). This increase compensates both the enhancement of the absorption rates induced by longer cavities and the shift of the inflection point of the input-intracavity non-linearity toward higher pump rates.

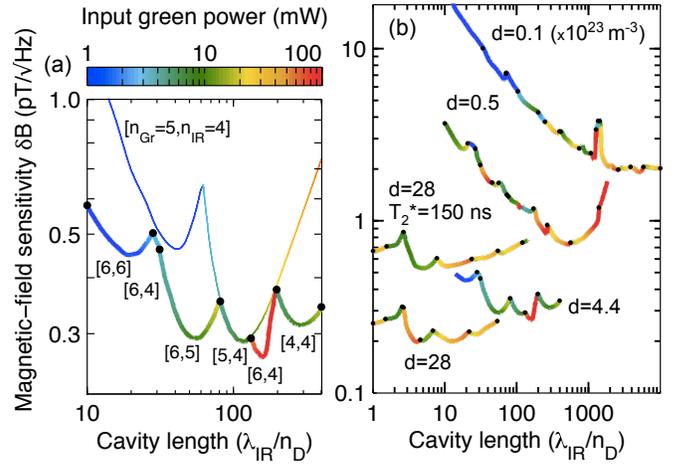

**Fig. 5.** (a,b) The bold lines indicate the optimized sensitivity to the magnetic field as the function of the cavity length. The color scale shows the green input power (arbitrarily limited to 100mW). The black dots indicate a change in the number of the green or infrared DBR pairs. (a) The density $d$ is equal to $4.4 \times 10^{23} \mathrm{m}^{-3}$ and $T_2^* = 390$ns. The thin line plots the sensitivity for $[n_{Gr} = 4, n_{IR} = 5]$. (b) $d$ is in $10^{23}\mathrm{m}^{-3}$ and $T_2^* = 390$ns (when it is not specified) or $T_2^* = 150$ns.

To find the best magnetic-field sensitivity that can be obtained for every cavity length, we simultaneously scan in Fig. 5a the magnetic-field sensitivity over the number of pairs on the two DBR pairs ($[N_{Gr}, N_{IR}] \in [\![0,8]\!]^2$, bold lines). The non-continuity of the curve derivative arises from discreet values of the mirrors reflectivity (given by their number of pairs). Curiously, short cavities lead to poorer sensitivities (factor $\sim 2$ between $10\lambda/n_D$ and $120\lambda/n_D$ long cavities) although mirrors with higher reflectivity could enhance the effective absorption length and restore the sensitivity. This observation means that the quality factor of the infrared cavity is not the only relevant parameter to describe the device. Other parameters such as the dependance of $\alpha_{NV}^s$ with the intracavity green and infrared powers – that depends on the overall losses – need to be considered.

The magnetic-field sensitivity also strongly depends on the density, $d$ of NV centers and on the electronic spin dephasing time, $T_2^*$ (Eq. 1). We first repeat the calculation for several $d$'s and find that higher densities of NV centers lead to a better sensitivity. Moreover, the best sensitivity case corresponds to shorter cavities and to lower green pump powers (Fig. 5b). This is because higher densities increase the absorption of the infrared light by the NV centers, $\alpha_{NV}^s$ then the contrast, $C$. Unfortunately, high densities may experimentally induced shorter coherence times $T_2^*$ [16]. Thus, we also calculate the sensitivity with $T_2^* = 150$ns when $d = 28 \times 10^{23}\mathrm{m}^{-3}$ [21, 30]. We observe that $\delta B$ is increased by a factor 2.6 when $T_2^*$ goes from 390ns to 150ns. This value is expected considering only the broadening of the ESR linewidth $\Gamma^{MW}$ in Eq. 2.

## 4. CONCLUSION

In summary, we have performed a systematic study on a two-color cavity system coupled via saturable absorbers. The best magnetic-field sensitivity that we obtain with real-world parameters corresponds to a shot-noise limited sensitivity as low as 290 fT/$\sqrt{\text{Hz}}$. This value is 520 (820) times lower than the configuration without the DBRs for transmission (reflection) measurements (Fig. 4 [31]).

We also observe and explain some of the non-linear effects induced by the saturable absorption lines of the NV centers and enhanced by the doubly resonant interacting cavities. This leads for instance to a strong dependance on the mirrors reflectivity and on the cavity length. Moreover, the magnetic-field sensitivity can, uncommonly, not be restored by increasing the mirrors reflectivity when decreasing the cavity length.

To even further improve the magnetic-field sensitivity, one could use a pulsed electron spin scheme that reduces the ESR linewidth [32]. The inherent scalability of the device will permit implementation of single-photonic chips containing arrays of such sensors to obtain one- or two- dimensional real-time images of the distribution of the magnetic field of samples.

### ACKNOWLEDGMENTS

We thank Thierry Debuisschert, Vincent Jacques and Glenn Solomon for fruitful discussions. This work was partially supported by the European Community's Seventh Framework Programme (FP7/2007-2013) under Grant Agreement No. 611143 (DIADEMS).

See Supplement 1 for supporting content.

# Supplemental Material: Highly sensitive on-chip magnetometer with saturable absorbers in two-color microcavities

## O. Gazzano[1,*,†] and C. Becher[1]

[1] *Fachrichtung 7.2 (Experimentalphysik), Universität des Saarlandes, Campus E2.6, 66123 Saarbrücken, Germany*
*Corresponding author: ogazzano@umd.edu*
[†] *Present address: Joint Quantum Institute, National Institute of Standards and Technology, & University of Maryland, Gaithersburg, MD, USA.*

---

## 1. DISTRIBUTION OF THE ELECTRIC FIELDS

We use a finite element software (COMSOL Multiphysics) to calculate the distribution of the electric field over the diamond waveguide including the DBR mirror structures for 532nm and 1042nm light fields without considering any absorption (Fig. S1). The calculation demonstrates the very good confinement of the two fields into the two-cavity system. Note that the amplitude of the green field is about 1.5 times stronger in the air layers of the infrared mirrors than in the cavity region due to the high refractive index of the diamond ($n_D \simeq 2.4$). This should not introduce extra losses for air/diamond DBRs because the absorption rates in the air is smaller than in the diamond layers that contain high densities of NV centers.

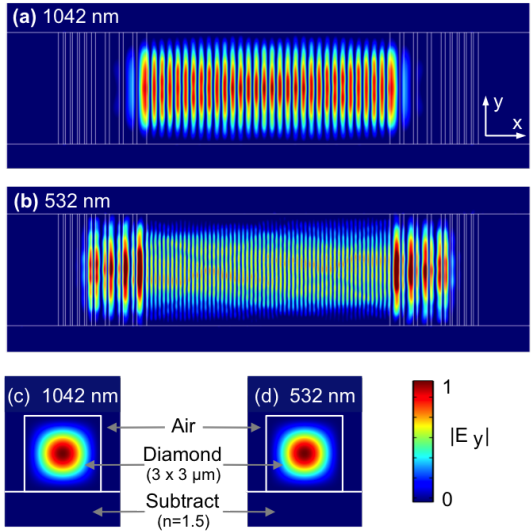

**Fig. S1.** Finite element calculations of the normalized modulus of the $y$ component of the electric field at 1042 nm (a,c) and 532 nm (b,d) with a $15\lambda/n_D$ long cavity, a $3 \times 3\mu m$ cross-section waveguide, surrounded with air and on a subtract with a refractive index $n = 1.5$. Calculations for a cut-section along x-y (a,b) and for a cut-section along y-z.

## 2. RATE EQUATION MODEL

We use the rate equation model described in [1] to calculate, in a stationary regime, the occupation probability $N_i$ of the 6 states $|i\rangle$ for $i = [\![1, 6]\!]$ of a single NV center. This model accounts for the transitions and the levels indicated in Fig. S2. We use the values indicated in Table S1 and in the main text to perform the calculations. In the low Rabi frequency $\Omega$ regime, the microwave transition rate is $W_{MW} = \Omega_2 T_2^* / 2$. We also consider $W_\lambda = \sigma_\lambda I_\lambda \lambda / (hc)$ where $I_\lambda$ in the intensity of the applied field.

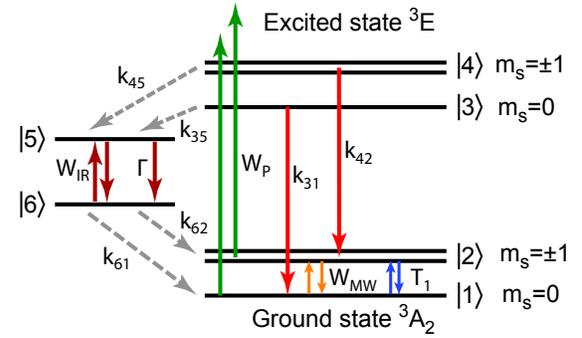

**Fig. S2.** Detailed energy levels and transition rates of the NV center – including the infrared transitions – considered for the calculations.

## 3. ESTIMATION OF THE MAGNETIC-FIELD SENSITIVITY

We use the rate equation model to calculate the absorption rates for the two fields and with the microwave field *on*- or *off*-resonance with the electronic spin transition:

$$\alpha_{NV}^{IR,s} = \sigma_{Gr}(N_1^s + N_2^s) \tag{1}$$

$$\alpha_{NV}^{Gr,s} = \sigma_{IR}(N_6^s - N_5^s) \tag{2}$$

In order to estimate the variations of the sensitivity with several parameters, we suppose in this section the unitarity of the input infrared beam power. Thus, the power after interaction with a single NV center is $P^s = 1 - \alpha_{NV}^{IR,s}$. The sensitivity can now be found using Eq. 1 of the main text. A scan over the green field allows us to extract the occupation probability of the state $|6\rangle$, the output power with the microwave field *on-* or *off-*resonance with the transition, as well as the magnetic-field sensitivity (Fig. S3a-c). The calculation for several values of infrared power reveals that the occupation probability $N_{|6\rangle}^{best}$ of the state $|6\rangle$ that corresponds to the best sensitivity decreases with the increase of the infrared pumping. Moreover the ratio of the occupation probability that gives the best sensitivity with the highest reachable occupation probability of the state $|6\rangle$, $N_{|6\rangle}^{best} / N_{|6\rangle}^{max}$ decreases from 0.7 at low field to 0.5 at high field.

| Parameter | Value | Unit | Ref. |
|---|---|---|---|
| $k_{31} = k_{42}$ | $66 \pm 5$ | $\mu s^{-1}$ | [2] |
| $k_{35}$ | $7.9 \pm 4.1$ | $\mu s^{-1}$ | [2] |
| $k_{45}$ | $53 \pm 7$ | $\mu s^{-1}$ | [2] |
| $k_{61}$ | $1.0 \pm 0.8$ | $\mu s^{-1}$ | [2] |
| $k_{62}$ | $0.7 \pm 0.5$ | $\mu s^{-1}$ | [2] |
| $\Gamma$ | 1 | $\mu s^{-1}$ | [3] |
| $\sigma_{Gr}$ | $3 \times 10^{-21}$ | $m^2$ | [4] |
| $\sigma_{IR}$ | $2 \times 10^{-22}$ | $m^2$ | [1] |

**Table S1.** Values of the transition rates indicated in Fig. S2 and of the green and infrared cross-sections.

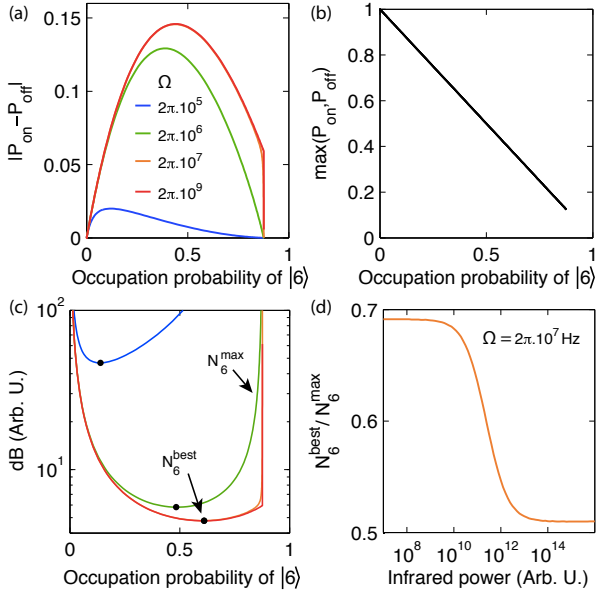

**Fig. S3.** (a-c) The following parameters are plotted as a function of the occupation probability of the state $|6\rangle$ with the microwave off-resonance: (a) $|P_{on} - P_{off}|$, (b) $\max((P_{on}, P_{off}))$ and (c) magnetic-field sensitivity deduced from (a) and (b). The microwave induces Rabi oscillations $\Omega$ indicated in Hz. The back dots mark the best sensitivity and correspond to $N_{|6\rangle}^{best}$.
(d) Ratio $N_{|6\rangle}^{best} / N_{|6\rangle}^{max}$ as a function of the infrared intensity for $\Omega = 2\pi \times 10^7 Hz$.

# 4. NUMERICAL CALCULATION OF THE SHOT-NOISE LIMITED SENSITIVITY

In order to determine the sensitivity to the magnetic field, we use a wave propagation model to calculate the intensity of the reflected and transmitted fields as a function of many different parameters of the system. To that aim, the medium is discretized in layers $p$ defined by a constant refractive index $n_p$. We also introduce sub-layers $q$ that are defined by the computational step $z_{step} = 10 nm$ of the calculation (Fig. A5). We write the electric field in a sub-layer by a vectorial written considering forward and backward propagation waves: $E^\lambda(q) = \left( E_F^\lambda(q), E_B^\lambda(q) \right)$ with $\lambda = \lambda_{Gr}$ or $\lambda_{IR}$. The transfer matrix from a sublayer $q$ to a sublayer $q - 1$ is diagonal [1]:

$$M_{q,q-1}^\lambda = \begin{pmatrix} M_{-1}^\lambda & 0 \\ 0 & M_{+1}^\lambda \end{pmatrix} \quad (3)$$

with

$$M_\epsilon^\lambda = \left[ 1 + \epsilon \frac{1}{2} (\alpha_0^\lambda + \alpha_{NV}^\lambda) \right] \cdot z_{step} \cdot e^{-\iota \cdot \epsilon \phi} \quad (4)$$

The term $\phi = 2\pi n_m z_{step} / \lambda$ is the dephasing within one computational sublayer and allows us to consider interfering effects between the forward and backwards propagating fields. The terms $\alpha_{NV}^\lambda$ depends on the intensity of both fields at the position of the NV center and on the microwave frequency. They are calculated at every computational steps using the equation rate model introduced above and in the main text. The transfer matrix between two layers $q$ and $q - 1$ of refractive index $n$ and $n - 1$ accounts only for the refractive index ratio and for the continuity of the electric field and of its derivative [5]:

$$M_{p,p-1} = \begin{pmatrix} 1 & 1 \\ 1 & -1 \end{pmatrix}^{-1} \begin{pmatrix} 1 & 1 \\ n_p / n_{p-1} & -n_p / n_{p-1} \end{pmatrix} \quad (5)$$

Since the green and infrared absorption rates by the NV centers non-linearly depend on both their intensities, we suppose that no light enters from the back mirror and we write $E_T^{Gr}$ and $E_T^{Ir}$ the amplitudes of the green and infrared light fields transmitted by the two-cavity system (Fig 5). Starting from the back of the structure, we calculate the amplitude of the input $E_{in}^\lambda$ and reflected $E_R^\lambda$ beams for the two colors (Fig. S4). The non-linear absorptions require us to calculate again and adjust both those transmitted values until the target values of $E_{in}^{Gr}$ and $E_{in}^{Ir}$ are reached [1]. The sensitivity to the magnetic field can now be calculated via Eq. 1 of the main text. Thereby, this model allows us to simulate the device for many different configurations.

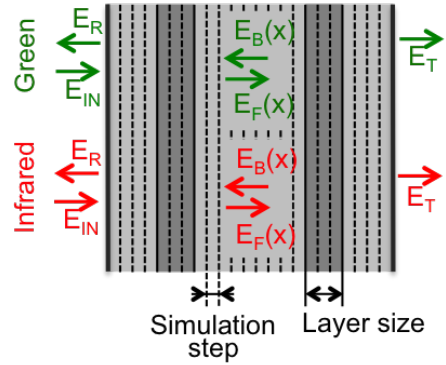

**Fig. S4.** Schematic of the model used for the simulations. The medium is divided in layers $p$ with a constant refractive index and sublayers $q$ for the computation. We consider forward and backward propagating fields for the two colors.

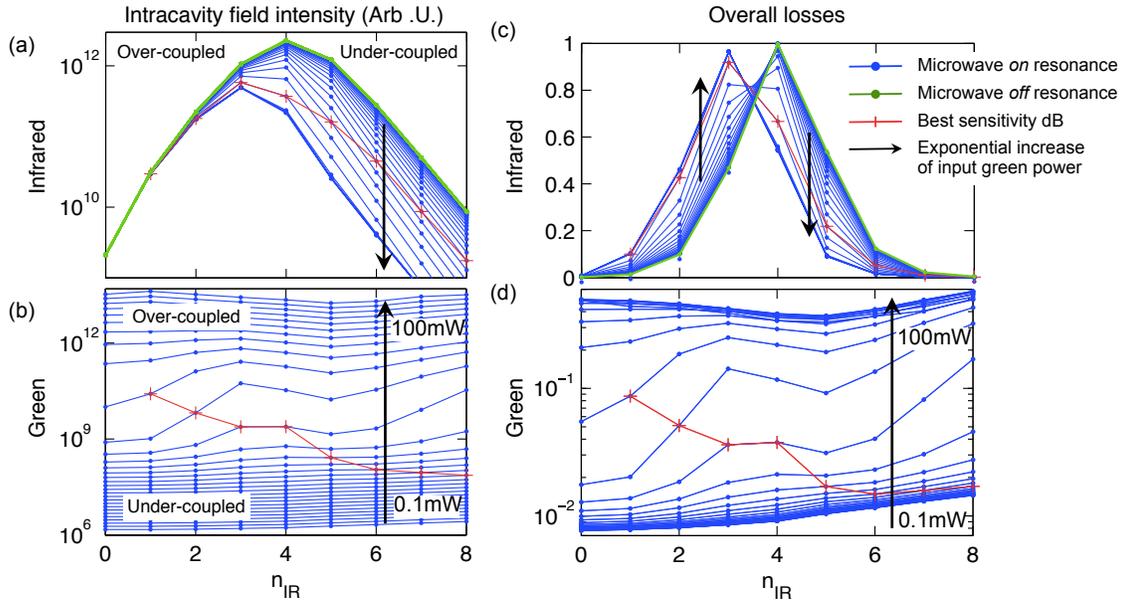

**Fig. S5.** For the infrared (a,c) and green (b,d) fields, we extract from the simulation runs: (a,b) the intracavity field intensity and (c,d) the overall losses. They are plotted for exponential increase of the green input power from 0.1mW to 100mW (arrows). The red curves indicate the green input power that corresponds to the best magnetic-field sensitivity. The microwave field is on- (blue) or off-resonance (red) with the transition. $n_{Gr} = 5$, $L = 120\lambda/n_d$, $n = 4.4 \times 10^{23}$ and $T_2^* = 390$ns.

## 5. TWO-COLOR CAVITIES COUPLING

To have a better understanding of the two cavity coupling, we extract from the simulation runs the intracavity field intensities for both the infrared and green lights. We consider the intracavity intensity fields right after the infrared mirror inside of the cavity. We clearly observe the transition between the over- to the under-coupled regime in Fig. S5a since the under-coupled cavity regime corresponds to a reduction of the intracavity field when the reflectivity of the mirror is increased. The overall losses, defined as $A^\lambda = 1 - (E_T^\lambda + E_R^\lambda)/R_{in}^\lambda$, for the infrared light and plotted in Fig. S5c are equal to one impedance matched regime. Interestingly, this case appears for different values of $n_{IR}$ when the green power changes due to a different equilibrium regime of the charges carries of the NV centers. The off-resonant microwave field corresponds to the lowest green pump power and this field is independent on the green power (green curves). We note that since the losses increase with the green power in the over-coupled regime and decrease in the under-coupled regime, a configuration where the losses with the microwave on- or off-resonant are equal exists and corresponds to the $F_R^{on} = F_R^{off}$ situation then to a configuration with $\delta B = +\infty$.

The figures S5b,d, that plot the intracavity green field intensity and the green overall losses, reveal the transition from the under coupled regime – where the pump is bellow the saturation regime of the NV centers – and the over coupled regime – where the field can be strongly enhanced in the cavity. Those figures show the coupling between the two cavities because the maximum of non-linearity – the impedance matching case – depends on the number of infrared pairs. Fig. S5b also shows that the maximum of sensitivity (red curves) corresponds to lower green intracavity field intensity when the reflectivity of the infrared mirrors increases. This is the sign of a change in the equilibrium regime of the NV center as shown in Fig. S3d.